\documentclass{mem}

\usepackage{natbib}
\usepackage{txfonts}
\usepackage{balance}
\usepackage{graphicx}
\usepackage[a4paper]{hyperref}
\idline{00}{111}

\def\eps@scaling{1.14}
\newcommand\plottwo[2]{{%
\typeout{Plottwo included the files #1 #2}
\centering
\leavevmode
\columnwidth=.9\columnwidth
\includegraphics[width={\eps@scaling\columnwidth}]{#1}%
\hfil
\hspace{-1.5cm}
\includegraphics[width={\eps@scaling\columnwidth}]{#2}%
}}%

\begin{document}

\title{Outskirts of spiral galaxies: result of a secular 
  evolution process?}

\subtitle{}

\author{J. \,Bakos\inst{1} \and I. \,Trujillo\inst{1} \and R.
 \,Azzollini\inst{1} \and J. E. \,Beckman\inst{1} \and
 M. \,Pohlen\inst{2}}

\offprints{Judith Bakos; \email{jbakos@iac.es}}
 
\institute{Instituto de Astrof\'isica de Canarias, La Laguna, Spain 
\and 
 School of Physics and Astronomy, Cardiff University, Cardiff, UK}

\authorrunning{Bakos et al.}

\titlerunning{Outskirts of spiral galaxies}

\abstract{We present our recent results on the properties of the
  outskirts of disk galaxies. In particular, we focus on spiral
  galaxies with stellar disk truncations in their radial surface
  brightness profiles. Using SDSS, UDF and GOODS data we show how the
  position of the break (i.e., a direct estimator of the size of the
  stellar disk) evolves with time since $z\sim1$. Our findings agree
  with an evolution on the radial position of the break by a factor of
  $1.3\pm0.1$ in the last 8 Gyr for galaxies with similar stellar
  masses. We also present radial color gradients and how they evolve
  with time. At all redshift we find a radial inside-out bluing
  reaching a minimum at the position of the break radius, this minimum
  is followed by a reddening outwards. Our results constrain several
  galaxy disk formation models and favour a scenario where stars are
  formed inside the break radius and are relocated in the outskirts of
  galaxies through secular processes.
 }

\maketitle{}

\begin{figure*}[t!]
\resizebox{\hsize}{!}{\includegraphics[clip=true]{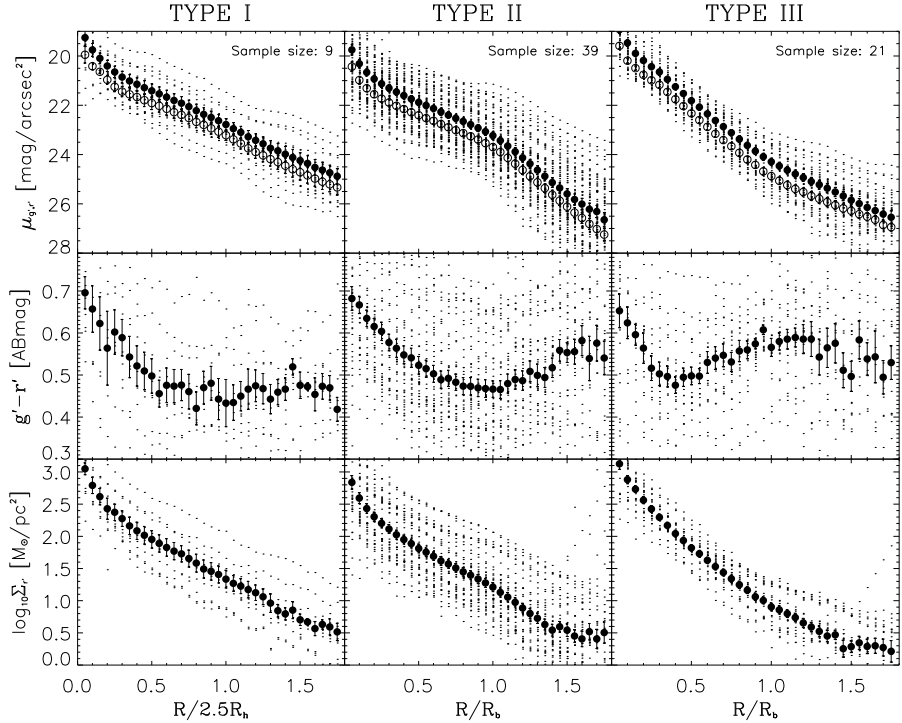}}
\caption{\footnotesize
  Upper panels: Averaged, scaled radial surface brightness profiles of
  9 Type~I (pure exponential profiles), 39 Type~II (truncated
  galaxies), and 21 Type~III (antitruncated) galaxies. The filled
  circles correspond to the $r'$-band mean surface brightness, the
  open circles to the mean $g'$-band data \citep{PT06}. The small dots
  are the individual galaxy profiles in both bands. The surface
  brightness is corrected for Galactic extinction. Middle panels: $(g'
  - r')$ color gradients. The averaged profile of Type~I reaches an
  asymptotic color value of $\sim 0.46$\,mag being rather constant
  outwards.  Type~II profiles have a minimum color of $0.47 \pm
  0.02$\,mag at the break position.  The mean color profile of
  Type~III has a redder value of about $0.57 \pm 0.02$\,mag at the
  break. Bottom panels: $r'$-band surface mass density profiles
  obtained using the color to $M/L$ conversion of \citet{Bell03}, and
  using Kroupa-IMF \citep{kroupa}. Note how the significance of the
  break almost disappears for the Type~II (truncated galaxies) case.}
\label{colorbak}
\end{figure*}

\section{Introduction}

Multiband observations \citep{gildepaz05, PT06, erwin08} show evidence
of a large number of stars being present in the outer regions of
spiral galaxy disks. However, current star formation theories do not
support easily the idea of stars forming in those regions, because the
environment is not dense enough to provide the conditions of star
formation, e.g., gas surface mass density is too low
($\le10$\,M$_{\odot}$\,pc$^{-2}$, \citealt{Kennicutt89}). This fact
creates a so-far not answered question: what is the origin of these
stars? We approach this problem by means of investigating the
structural and stellar population properties of the (outskirts) spiral
galaxy disks.

Early studies of the disks of spiral galaxies \citep{Patterson40,
vauc58, freeman70} showed that this component generally follows an
exponential radial surface-brightness profile, with a certain
scale-length, usually taken as the characteristic size of the
disk. \citet{freeman70} pointed out, though, that not all disks follow
this simple exponential law. In fact, a repeatedly reported feature of
disks for a representative fraction of the spiral galaxies is that of
a truncation \citep{Kruit79} of the stellar population at large radii,
typically 2--4 exponential scale-lengths \citep[see e.g., the review
by][]{Pohlen04}.

\begin{figure*}[t!]
\plottwo{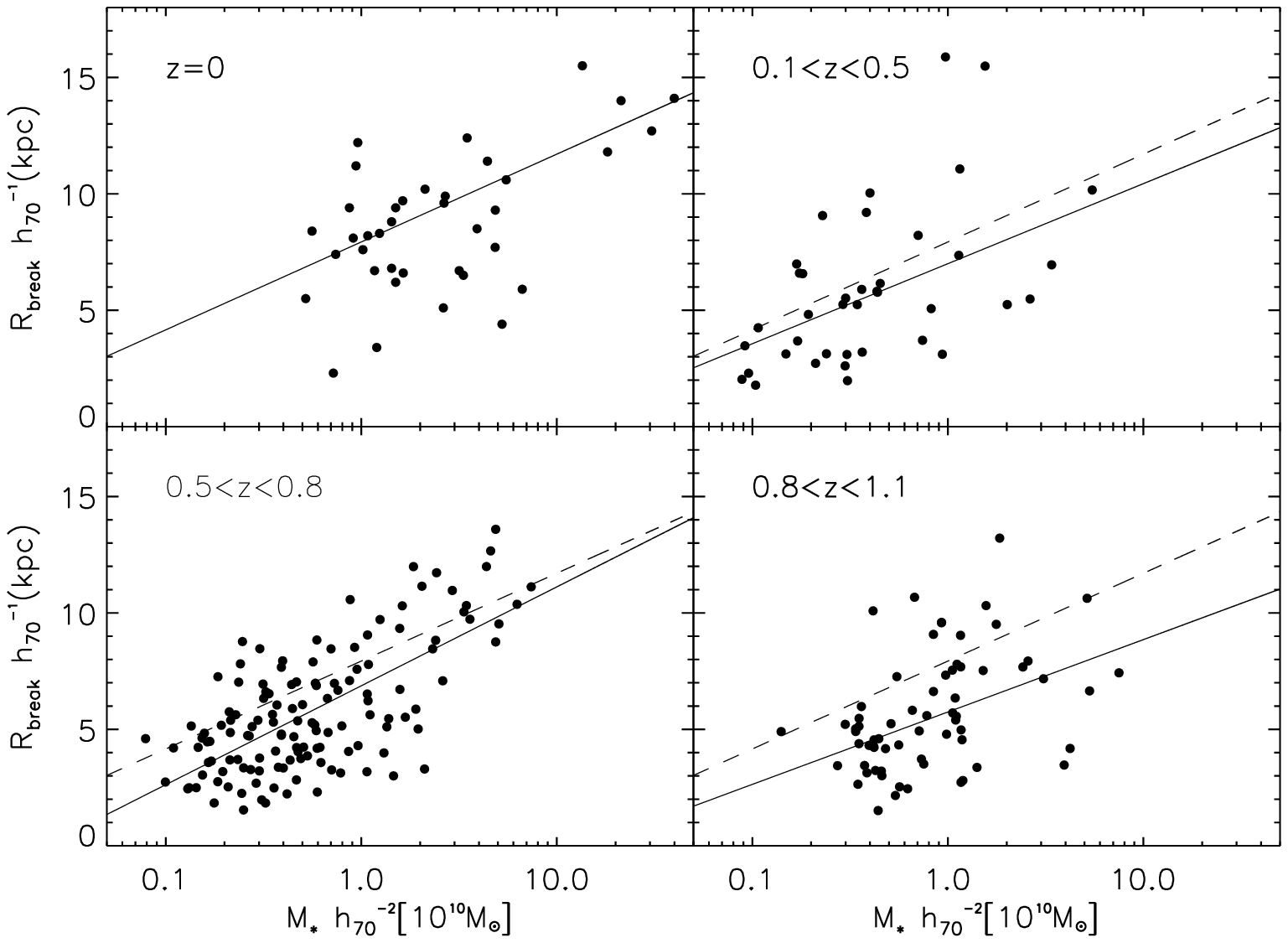}{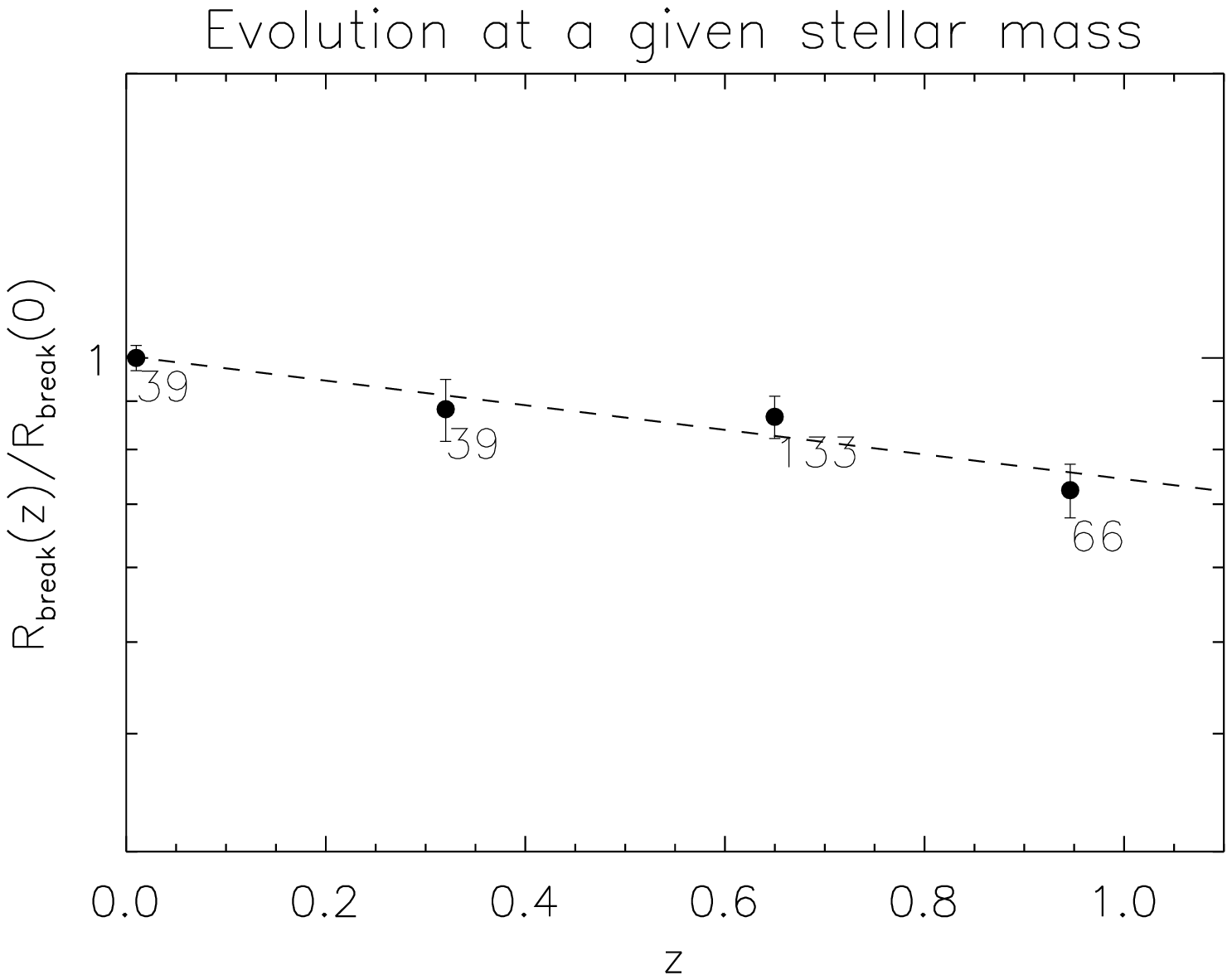}
\caption{\footnotesize
  Left: Break radius of `truncated' galaxies as a function of
  stellar mass, for 4 ranges of redshift. Local data are from
  \citet{PT06} ($g'$-band results). Right: Size evolution at a given
  stellar mass of the break radius as a function of redshift. We have
  found a growth of a factor $1.3\pm0.1$ between $z=1$ and $z=0$. The
  numbers below each point in the right panel indicate the size of the
  sample.}
\label{figTrmass}
\end{figure*}

Several possible break-forming mechanisms have been investigated to
explain the truncations. There have been ideas based on maximum
angular momentum distribution: \citet{Kruit87} proposed that angular
momentum conservation in a collapsing, uniformly rotating cloud
naturally gives rise to disk breaks at roughly 4.5 scale
radii. \citet{Bosch01} suggested that the breaks are due to angular
momentum cut-offs of the cooled gas. On the other hand, breaks have
also been attributed to a threshold for star formation, due to
changes in the gas density \citep{Kennicutt89}, or to an absence of
equilibrium in the cool interstellar medium phase
\citep{ElmegreenParravano94, Schaye04}. Magnetic fields have been also
considered \citep{Battaner2002} as responsible of the
truncations. More recent models using collisionless N-body
simulations, such as those by \citet{Debattista06}, demonstrated that
the redistribution of angular momentum by spirals during bar formation
also produces realistic breaks. In a further elaboration of this idea,
\citet{Roskar08} have performed high resolution simulations of the
formation of a galaxy embedded in a dark matter halo. In these models,
breaks are the result of the interplay between a radial star formation
cut-off and redistribution of stellar mass by secular processes. A
natural prediction of these models is that the stellar populations
present an age minimum in the break position. This prediction could be
probed by exploring the color profiles of the galaxies.

Furthermore, addressing the question of how the radial truncation
evolves with $z$ is strongly linked to our understanding of how the
galactic disks grow and where star formation takes
place. \citet{Perez04} showed that it is possible to detect stellar
truncations even out to $z\sim1$. Using the radial position of the
truncation as a direct estimator of the size of the stellar disk,
\citet{TP05} inferred a moderate ($\sim 25$\%) inside-out growth of
disk galaxies since $z\sim1$. An important point, however was missing
in the previous analyses: the evolution with redshift of the radial
position of the break at a given stellar mass. The stellar mass is a
much better parameter to explore the growth of galaxies, since the
luminosity evolution of the stellar populations can mimic a size
evolution \citep{Trujillo04, Trujillo06}. We present in this
contribution a quick summary of our recent findings on the stellar
disk truncation origin and its evolution with redshift. The results
presented here are based on the following publications:
\citet{Azzollini08a, Azzollini08b} and \citet{Bakos08}. 
Throughout, we assume a flat $\Lambda$-dominated cosmology
($\Omega_{\rm m} = 0.30$, $\Omega_{\Lambda} = 0.70$, and $H_{0} = 70$
km\,s$^{-1}$\,Mpc$^{-1}$).

\section{Color profiles of local galaxies}

In order to constrain the outer disk formation models, in
\citet{Bakos08}, we have explored radial color and stellar surface
mass density profiles for a sample of 85 late-type spiral galaxies
with available deep (down to $\sim27$ mag\,arcsec$^{-2}$) SDSS $g'$-
and $r'$-band surface brightness profiles \citep{PT06}. About $90$\%
of the light profiles have been classified as broken exponentials,
exhibiting either truncations (Type~II galaxies) or antitruncations
(Type~III galaxies). Their associated color profiles show a
significantly different behavior. For the truncated galaxies a radial
inside-out bluing reaches a minimum of $(g'-r') = 0.47 \pm 0.02$ mag
at the position of the break radius, this minimum is followed by a
reddening outwards (see the middle panels in Fig.~\ref{colorbak}). The
antitruncated galaxies reveal a different behavior. At the position of
the break radius (obtained from the light profiles) now resides a
plateau region of the color profile with a value about $(g'-r') = 0.57
\pm 0.02$.

\subsection{Stellar surface mass density profiles}

Using the $(g'-r')$ color \citep{Bell03} it is possible calculate
mass-to-light ($M/L$) ratios along the radius of the disks. Converting
the $(M/L)$ into stellar surface mass density reveals a surprising
result. The breaks, well established in the light profiles of the
Type~II galaxies, are almost gone (in case of several individual
galaxies, e.g., NGC~5300, the break is completely gone). The mass
profiles resembles now those of the pure exponential Type~I galaxies
(see the bottom panels in Fig.~\ref{colorbak}). This result suggests
that the origin of the break in Type~II galaxies is more likely due to
a radial change in the ingredients of the stellar population than
being associated to an actual drop in the distribution of mass. The
antitruncated galaxies, on the other hand, show clear mass-excess in
the outer regions on the stellar mass density profiles, which could
have been accumulated from an external (possibly satellite) origin.

There are other structural parameters that can be computed to constrain
the different formation scenarios. Among these we have estimated the
stellar surface mass density at the break for truncated (Type~II)
galaxies ($13.6\pm1.6$ M$_{\odot}$\,pc$^{-2}$) and the same parameter
for the antitruncated (Type~III) galaxies ($9.9\pm1.3$
M$_{\odot}$\,pc$^{-2}$). Finally, we have measured that \mbox{$\sim15$\%} of
the total stellar mass in case of truncated galaxies and $\sim9$\% in
case of antitruncated galaxies are to be found beyond the measured
break radii in the light profiles.

\section{Stellar disk truncation along the Hubble-time}

\begin{figure}[]
\resizebox{\hsize}{!}{\includegraphics[clip=true]{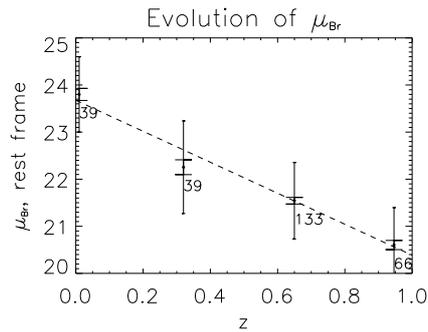}}
\caption{\footnotesize 
  Evolution of the surface brightness at the break for Type II
  galaxies with redshift. We show the median surface brightness at the
  break for the distribution of our galaxies. The larger error bars
  represent the standard deviation of the distributions, while the
  shorter ones give the error in the median values. (The numbers below
  each point give the size of the sample at that redshift bin.)}
\label{figmu}
\end{figure}

In \citet{Azzollini08a}, we have conducted the largest systematic
search so far for stellar disk truncations in disk-like galaxies at
intermediate redshift ($z<1.1$), using the Great Observatories Origins
Deep Survey South (GOODS-S) data from the Hubble Space
Telescope/ACS. Focusing on Type II galaxies (i.e., downbending
profiles) we explore whether the position of the break in the
rest-frame $B$-band radial surface brightness profile (a direct
estimator of the extent of the disk where most of the massive star
formation is taking place), evolves with time. The number of galaxies
under analysis (238 of a total of 505) is an order of magnitude larger
than in previous studies. For the first time, we probe the evolution
of the break radius for a given stellar mass (a parameter well suited
to address evolutionary studies). Our results suggest that, for a
given stellar mass, the radial position of the break has increased
with cosmic time by a factor $1.3\pm0.1$ between $z\sim1$ and $z\sim0$
(see Fig.~\ref{figTrmass}). This is in agreement with a moderate
inside-out growth of the disk galaxies in the last $\sim8$ Gyr. In the
same period of time, the surface brightness level in the rest-frame
$B$-band at which the break takes place has increased by $3.3\pm0.2$
mag\,arcsec$^{-2}$ (a decrease in brightness by a factor of
$20.9\pm4.2$, see Fig.~\ref{figmu}).

\begin{figure*}[t!]
\resizebox{\hsize}{!}{\includegraphics[clip=true]{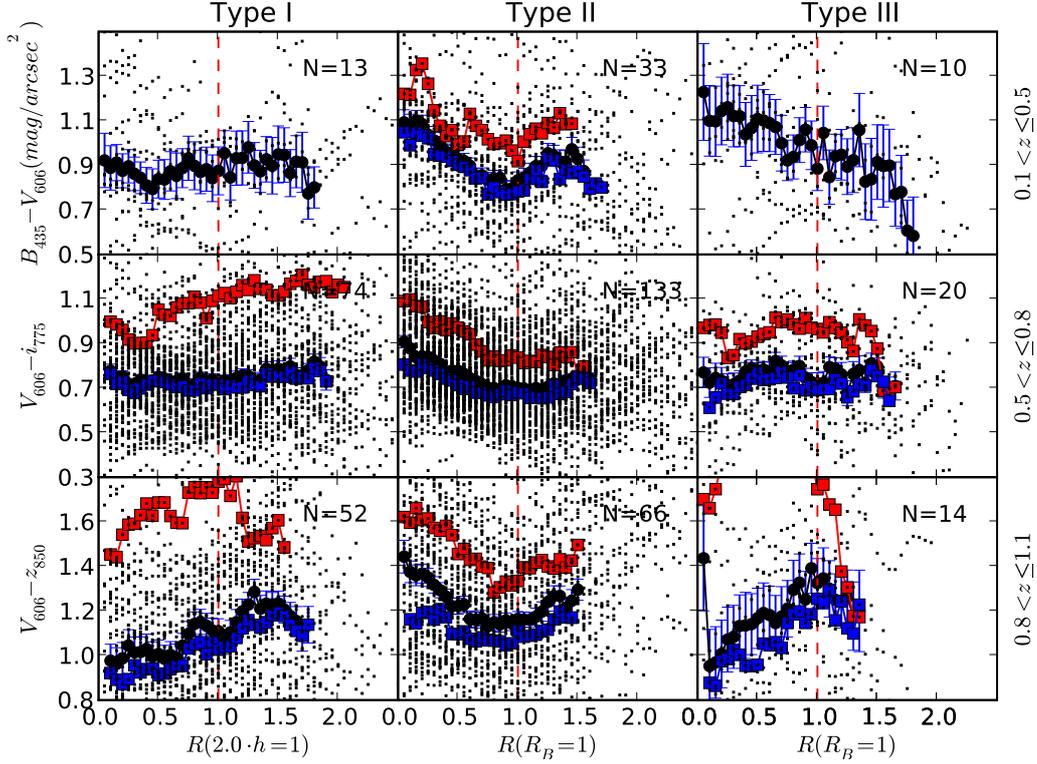}}
\caption{\footnotesize
 Color profiles of the 415 galaxies under study in
 \citet{Azzollini08b}. The sample is divided in subsamples according
 to surface brightness profile type (I or pure exponential profiles,
 II or truncated galaxies, III or antitruncated, in columns, from left
 to right) and redshift range (low, mid, or high, in rows, from top to
 bottom). The colors ($B_{435}-V_{606}$, $V_{606}-i_{775}$,
 $V_{606}-z_{850}$) are chosen as the best proxies to the rest-frame
 $u-g$ color in each redshift bin. The radii are scaled to the scale
 radius, $R_{\rm s}$, whose definition depends on profile type:
 $R_{\rm s}=2h$ for type I, where $h$ is the scale-length of the disk,
 and it is equal to the break radius, $R_{\rm s}=R_{\rm B}$, for types
 II and III. Small points are individual color profiles. Large black
 dots are the median color profiles for each subsample, and the error
 bars give the error in those estimations. The red squares give the
 median color profile for objects with stellar mass
 $M_{\star}>10^{10}$ M$_{\odot}$, while the blue squares give the same
 for objects with $M_{\star}\leq10^{10}$ M$_{\odot}$.}
\label{colorazzo}
\end{figure*}

In \citet{Azzollini08a} we also find that at a given stellar mass, the
scale-lengths of the disk in the part inner to the `break' were on
average somewhat larger in the past, and have remained more or less
constant until recently. This phenomenon could be related to the
spatial distribution of star formation, which seems to be rather
spread over the disks in the images. So disk galaxies had profiles
with a flatter brightness distribution in the inner part of the disk,
which has grown in extension, while becoming fainter and `steeper'
over time. This is consistent with at least some versions of the
inside-out formation scenario for disks.

\section{Color profiles in intermediate redshift galaxies}

In addition to the evolution of the position of the break in spiral
galaxies, it is important to explore how the color of the surface
brightness profiles has evolved with time. This kind of analysis sheds
light on when stars formed in different parts of the disk of galaxies,
thus giving hints on the stellar mass buildup process.

In \citet{Azzollini08b} we present deep color profiles for a sample of
415 disk galaxies within the redshift range $0.1 \leq z \leq 1.1$, and
contained in HST/ACS imaging of the GOODS-South field. For each
galaxy, passband combinations are chosen to obtain, at each redshift,
the best possible approximation to the rest-frame $u-g$ color (see,
Fig.~\ref{colorazzo}). We find that objects which show a truncation in
their stellar disk (type II objects) usually show a minimum in their
color profile at the break, or very near to it, with a maximum to
minimum amplitude in color of $\leq0.2$ mag\,arcsec$^{-2}$, a feature
which is persistent through the explored range of redshifts (i.e., in
the last $\sim8$ Gyr and that it is also found in our local sample for
comparison \citep{Bakos08}. This color structure is in qualitative
agreement with recent model expectations where the break of the
surface brightness profiles is the result of the interplay between a
radial star formation cutoff and a redistribution of stellar mass by
secular processes \citep{Roskar08}.

\section{Discussion}

Our results on the color profiles fit qualitatively with the
particular prediction of \citet{Roskar08}, that the youngest stellar
population should be found at the break radius, and older (redder)
stars must be located beyond that radius. It is not easy to understand
how `angular momentum' or `star formation threshold'/`ISM phases'
models alone could explain our results. Thus they pose a difficult
challenge for these models. However, it will also be necessary to
check whether the \citet{Roskar08} models (as well as other available
models in the literature like those of \citet{Bour07} and
\citet{Foyle08}) are able to reproduce quantitatively the results shown
here.

Combining the results found in \citet{Azzollini08b} and
\citet{Bakos08} one is tempted to claim that both the existence of the
break in Type~II galaxies, as well as the shape of their color
profiles, are long lived features in the galaxy evolution. Because it
would be hard to imagine how the above features could be continuously
destroyed and re-created maintaining the same properties over the last
$\sim8$ Gyr.

%%% BIBLIOGRAPHY

\bibliographystyle{aa}

\begin{thebibliography}{}

\bibitem[Azzollini et al.(2008a)]{Azzollini08b} 
  Azzollini, R., Trujillo, I., \& Beckman, J. E.\ 2008a, \apj, 679, L69

\bibitem[Azzollini et al.(2008b)]{Azzollini08a} 
  Azzollini, R., Trujillo, I., \& Beckman, J. E.\ 2008b, \apj, 684, 1026

\bibitem[Bakos et al.(2008)]{Bakos08} 
  Bakos, J., Trujillo, I., \& Pohlen, M. \ 2008, \apj, 683, L103

\bibitem[Battaner et al.(2002)]{Battaner2002} 
  Battaner, E., Florido, E., \& Sanchez-Saavedra, M.~L.\ 1992, \aap, 253, 89 

\bibitem[Bell et al.(2003)]{Bell03} 
  Bell, E. F., McIntosh, D. H., Katz, N., \& Weinberg, M. D. \ 2003,
  \apjs, 149, 289

\bibitem[Bournaud et al.(2007)]{Bour07} 
  Bournaud, F., Elmegreen, B. G., \& Elmegreen, D. M. \ 2007, \apj,
  670, 237

\bibitem[Debattista et al.(2006)]{Debattista06} 
  Debattista, V. P., Mayer, L., Carollo, C. M., et al. \ 2006, \apj,
  645, 209

\bibitem[de Vaucouleurs(1958)]{vauc58} 
  de Vaucouleurs, G. \ 1958, \apj, 128, 465

\bibitem[Elmegreen \& Parravano(1994)]{ElmegreenParravano94} 
  Elmegreen, B. G., \& Parravano, A. \ 1994, \apj, 435, L121

\bibitem[Erwin et al.(2008)]{erwin08} 
  Erwin, P., Pohlen, M., \& Beckman, J. E. \ 2008, \aj, 135, 20

\bibitem[Foyle et al.(2008)]{Foyle08} 
  Foyle, K., Courteau, S., \& Thacker, R. J. \ 2008, \mnras, 386, 1821

\bibitem[Freeman(1970)]{freeman70} 
  Freeman, K. C. \ 1970, \apj, 160, 811

\bibitem[Gil De Paz et al.(2005)]{gildepaz05} 
  Gil de Paz, A., Madore, B. F., Boissier, S. et al. \ 2005, \apj,
  627, L29

\bibitem[Kennicutt(1989)]{Kennicutt89} 
  Kennicutt, R. C. \ 1989, \apj, 344, 685

\bibitem[Kroupa(2001)]{kroupa} 
  Kroupa, P. \ 2001, \mnras, 322, 231

\bibitem[Patterson(1940)]{Patterson40} 
  Patterson, F. S. 1940, Harvard Coll. Obs. Bul., 914, 9

\bibitem[P\'erez(2004)]{Perez04} 
  P\'erez, I. \ 2004, \aap, 427, L17


\bibitem[Pohlen \& Trujillo(2006)]{PT06} 
  Pohlen, M., \& Trujillo, I. \ 2006, \aap, 454, 759

\bibitem[Pohlen et al.(2004)]{Pohlen04}
  Pohlen, M., Beckman, J. E., H\"uttemeister, S., et al. 2004, in
  Penetrating Bars through Masks of Cosmic Dust, ed. D. L. Block,
  I. Puerari, K. C. Freeman, et al. (Springer, Dordrecht), 731

\bibitem[Ro\v skar et al.(2008)]{Roskar08} 
  Ro\v skar, R., Debattista, V. P., Stinson, G. S., et al. \ 2008,
  \apj, 675, L65

\bibitem[Schaye(2004)]{Schaye04} 
  Schaye, J. \ 2004, \apj, 609, 667

\bibitem[Trujillo \& Pohlen(2005)]{TP05} 
  Trujillo, I., \& Pohlen, M. \ 2005, \apj, 630, L17

\bibitem[Trujillo et al.(2004)]{Trujillo04} 
  Trujillo, I., Rudnick, G., Rix, H.-W., et al. \ 2004, \apj, 604, 521

\bibitem[Trujillo et al.(2006)]{Trujillo06} 
  Trujillo, I., F\"orster Schreiber, N. M., Rudnick, G., et al. \
  2006, \apj, 650, 18

\bibitem[van den Bosch(2001)]{Bosch01} 
  van den Bosch, F. C. \ 2001, \mnras, 327, 1334

\bibitem[van der Kruit(1979)]{Kruit79} 
  van der Kruit, P. C. \ 1979, \aaps, 38, 15

\bibitem[van der Kruit(1987)]{Kruit87} 
  van der Kruit, P. C. \ 1987, \aap, 173, 59

\end{thebibliography}

\end{document}